\documentclass{amsart}

\usepackage{natbib}
\numberwithin{equation}{section}

\usepackage{graphics}
\usepackage[dvipdf]{graphicx}

\usepackage{epsfig}
\usepackage{url}

\usepackage{epsfig}

\usepackage{amssymb}

\usepackage{pifont}

\usepackage{supertabular}

\begin{document}

\title{Failure modes of complex materials with spatially-correlated mechanical properties - the critical role of internal damage}
\author{J.~Faillettaz}
\address{STEP, ETH Z\"urich, 8092 Z\"urich, Switzerland}
\curraddr{3G, University of Zurich, 8057 Zurich, Switzerland
}

\author{D.~Or} 
\address{STEP, ETH Z\"urich, 8092 Z\"urich, Switzerland}
\begin{abstract}

The study reports a systematic evaluation of the role of spatially correlated mechanical elements on failure behavior of heterogeneous materials represented by fiber bundle models (FBM) with different load redistribution rules.
The increase of spatial correlation FBM for a local load sharing, results in a transition from ductile-like failure characteristics into brittle-like failure. The study identified a global failure criterion based on macroscopic properties (external load and cumulative damage) which is independent of spatial correlation or load redistribution rules. This invariant metric could be applied for early warning of a class of geophysical ruptures. 

\end{abstract}


\maketitle

Various geomorphological processes and natural hazards are triggered by gravity driven instabilities (landslides, rockfalls, snow avalanches, and glacier failure). Common to such  phenomena is the abrupt and often extensive failures within heterogeneous natural materials. In the context of natural hazards, the prediction of such catastrophic events is required for developing effective mitigation measures. Unfortunately, the predictability of such events is hindered by the complex and highly non-linear failure processes within inherently heterogeneous media.

Certain brittle geologic media may fail by propagation of a single extensive crack, and considerable efforts have been devoted to the understanding, detection, and prevention of crack nucleation \cite{Fineberg&Marder1999}. The failure mode of heterogeneous materials is often more gradual and ductile \cite{Sornette2006}. Catastrophic failure of heterogeneous materials occurs as the culmination of a progressive damage involving complex interactions between multiple defects and growing micro-cracks \cite{Sornette2006}. These precursor internal damage events that may be detected through their acoustic emission \cite{Michlmayr&al2012} as has been shown for rockfall \citep{Amitrano&al2005} or soil slope failure \cite{Dixon&al2003,Kolesnikov&al2003,Dixon&Spriggs2007}.
Models for mechanical interactions among many elements provide a means for studying progression towards failure, including the branching model, the thermal fuse model, the spring- block model and the fiber bundle model (see \cite{Sornette2006,Alava&al2006} for a review). The flexibility and simplicity of the fiber bundle model (FBM) offer a particularly useful framework for systematically studying processes preceding global failure \cite{Gomez&al1993,Kloster&al1997,Pradhan&al2010}. Additionally, the discrete nature of failure events offers a direct link with acoustic emissions suggested for monitoring such progressive failures \cite{Michlmayr&al2012}.
The mechanical properties of natural materials are characterized by structural spatial correlation resulting from the combined action of several physical, chemical or biological processes and from complex interactions between geology, topography, climate or soil use \cite{Tangmar&al1985}.
However, effect of such spatial correlations on the rupture maturation process has not been studied so far.  

This study systematically evaluates the influence of spatial organisation of mechanical properties on rupture behavior and its implications for precursory events linked with early warning for natural hazard prediction.   
Following a brief description of the different numerical models developed in this work, the rupture behavior and associated macroscopic properties at rupture of simulated FBMs are presented for different spatial arrangements. 
Precursory activity preceding complete breakdown is also examined in a statistical framework aimed at developing metrics for failure imminence. 
Finally, a general rupture criterion based on macroscopic properties is presented, and its possible application to gravity-driven geophysical instabilities is discussed.

{\it Failure Events in Correlated Fiber Bundle Model.}-
We consider a set of parallel elasto-brittle fibers assembled on a square regular lattice with periodical boundary conditions. 
The fibers exhibit a linear-elastic behavior (with a Young modulus $E$ [Pa] arbitrarily set to a value E=1) followed by an abrupt failure at a prescribed critical threshold load termed $\sigma_{th}$ (values drawn from a prescribed probability distribution, i.e., uniform, Weibull or lognormal distribution).
The load carried by a failed fiber is subsequently redistributed equally to surviving fibers by either a global ("democratic") load sharing rule (denoted as DFBM) or by local load sharing to nearest neighbors. 
We considered two modes of local load sharing: the nearest 4 neighbors on a square lattice (denoted as LFBM4) or eight nearest neighbors (denoted as LFBM8).

\begin{figure}
\includegraphics[width=8.6cm]{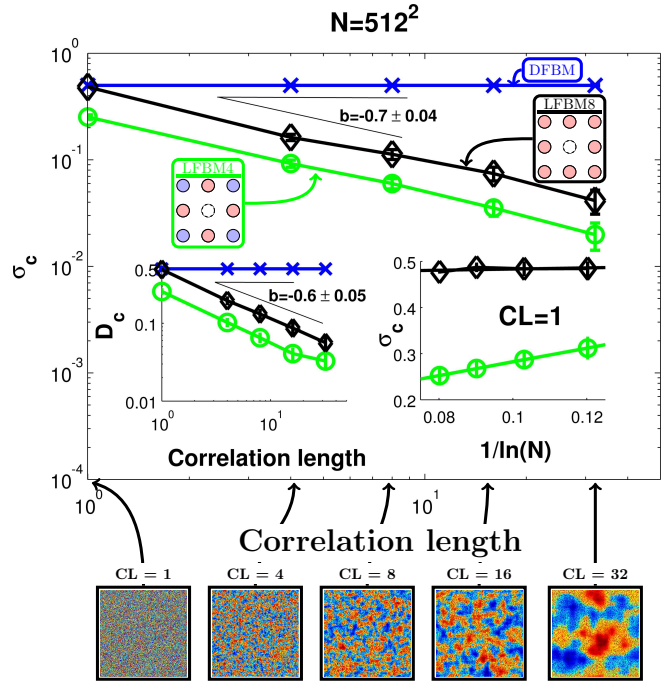}
\caption{\label{sigmac} Critical effective stress $\sigma_c$ (inset left: damage $D_c$) at rupture as a function of correlation length for DFBM (blue), LFBM4 (green) and LFBM8 (black). 
Inset right: Critical effective stress as a function of domain size $N$ for CL=1 for LFBM4 (green) and LFBM8 (black). The bottom raw of insets shows the relative organization of the strength (blue: weak fiber, red: strong) with different CL.}
\end{figure}

The effects of heterogeneity patterns on the global rupture behavior of the bundle were studied using the values of fiber strengths $\sigma_{th}$ chosen randomly from a uniform distribution in the range from 0 to 1. For robust statistical representation we generated 100 different initial configurations of each FBM studied.
Moreover, to account for natural spatial correlations (such as induced by tree root system with large lateral extent), we reorganized the initially uniformly distributed configuration of fiber strengths to obtain different spatial correlation lengths (see Fig. \ref{sigmac} insets).
The process involved three steps.
 First, an optimized correlated lattice with a given correlation length CL (expressed in terms of lattice units) was computed by convolving in the Fourier space a Gaussian filter with a lattice composed of uniformly distributed fiber strengths. 
 Second, each fiber of the initial lattice was spatially migrated to the position of the fiber in the optimized lattice with the nearest strength value.
 Finally, a test on the reorganized lattice was performed to verify that the fiber strengths $\sigma_{th}$ were organized according to a given correlation length (Fig. \ref{sigmac}). 
 It is important to note that both initial and reorganized lattice were composed of the same number of fibers with the same values of fiber strengths, only their spatial organization differs. 
The different FBMs were loaded by matching each load step to the weakest intact fiber (an adaptive load-controlled mode). This loading mode is closest to quasistatic gravity driven instabilities similar to gradual loading of a hillslope by rainfall. 
Different simulations were performed with different lattice size ranging from $N=64^2$ to $512^2$ fibers, with different stress redistributions (democratic sharing - DFBM, local sharing with coordination number of 4 - LFBM4 or 8 - LFBM8) and with different initial correlation length {\it CL} (ranging from 1 to 32 lattice units).

{\it Results.}-
The effect of spatial correlation of the same mechanical elements on the effective critical stress $\sigma_c$ (the total force on intact fibers divided by their number) and the fraction of failed fibers (denoted as FBM damage, $D_c$) for different stress distribution rules are depicted in Fig. \ref{sigmac}. 
As expected, the correlation length (CL) exerted no influence on the rupture occurrence in global load sharing (DFBM). For the local load sharing scenario (LFBM), increasing the range of CL resulted in a decrease in the critical effective stress, $\sigma_c$. 
Note that the DFBM redistributes the load of failed fibers to all the intact fibers regardless of their positions, hence, spatial organization of fiber strengths has no influence on the behavior of global failure of the bundle. The failure behavior of DFBM is  progressive and proceed in a ductile-like fashion.
In contrast, for local load sharing where the redistributed load is concentrated in the neighborhood of a broken fiber, the presence of spatial correlation length (CL) changes drastically failure behavior. The probability of neighboring fibers with similar strengths increases with increasing CL, hence the redistributed load is more likely to trigger a cascade of failures of near neighbors. 
Consequently failure of groups of fibers (termed "avalanches") are also promoted and are likely to cascade into global failure in a brittle-like fashion.
Note that DFBM and LFBM8 exhibit similar critical load $\sigma_c$ for the shortest correlation length of CL=1 whereas global failure occurs at about half the load for LFBM4.
The critical load $\sigma_c$ at global failure depends on the coordination number - $cn$ ($cn$=4 or 8 neighbors) as well as on correlation length - $CL$ (Fig \ref{sigmac}) and can be expressed (for a square lattice) as: $\sigma_c(cn, CL) \sim \sigma_c(cn,1) \times CL^{-0.7}$.
Cumulative damage $D_c$ at global failure (i.e., fraction of broken fibers in inset of Fig \ref{sigmac}) decreases with increasing CL following approximately the same power-law behavior as $\sigma_c$ 
(i.e, $D(cn, CL) \sim D(cn,1) \times CL^{-0.6}$). 
Interestingly, for large CL, the onset of global failure occurs with very low damage (4 $\%$) or fraction of failed fibers (but not their strength, as we shall see next). Consequently, not many precursory signals for impending global failure based on microseismic or acoustic emissions \cite{Michlmayr&al2012} activity that are likely to be  generated in a heterogeneous slope with a large CL.
Note also that DFBM and LFBM8 for CL = 1 have the same damage at global failure.
The LFBM8 with CL=1 exhibits a similar macroscopic behavior as DFBM (where rupture proceeds in a ductile-like fashion) however, when only CL increases, the behavior shifts into a brittle-like rupture similar to LFBM4: In other words, the presence of spatial correlation in mechanical properties affects the nature of events preceding global failure, shifting from progressive diffuse damage (ductile-like rupture) for uniformly random distribution with CL=1 to abrupt single crack growth (brittle-like rupture) with increasing CL.
The stress redistribution localization appears also to play an important role on global failure behavior. The right bottom inset of Fig. \ref{sigmac} shows the dependence of the total number of fibers in the LFBM domain (or domain size) on the value of $\sigma_c$. It appears that $\sigma_c \sim 1/ln~N$ (where $N$ is the total number of fibers) for LFBM4 as \cite{Hidalgo&al2002} already pointed out. Moreover, no dependency was found for LFBM8, suggesting that the crossover between short and long range interactions \cite{Kun&al2003} may have "saturated" for the eight neighbors.

\begin{figure}
\includegraphics[width=8.6cm]{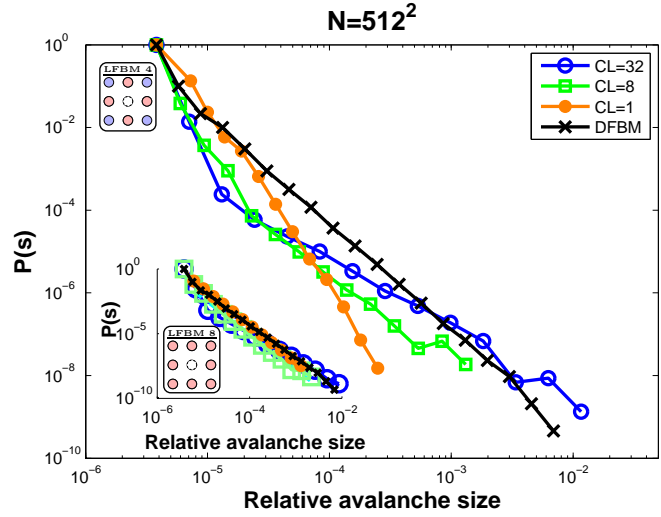}
\caption{\label{SFD} Avalanche size-frequency distribution of 100 runs for different correlation lengths for LFBM4  and LFBM8 (see inset for LFBM8). }
\end{figure}

The failure process can also be characterized by the precursory activity prior to global failure. The statistical properties of fiber failure sequences are characterized by the size distribution of groups of fibers that fail simultaneously ("avalanche") that, from an experimental point of view, could be related to the amplitudes of acoustic emissions generated during the gradual failure of materials \cite{Garcimartin&al1997}.
The avalanche size-frequency distributions (SFD) for the different FBM are depicted in Fig. \ref{SFD}.
The SFD appears to exhibit a power-law behavior for all FBM for a correlation length of 1 but with different power-law exponents (termed here $b$-exponent \cite{Amitrano2012}), ranging from -2.5 for DFBM (which conforms with analytical results of \cite{Alava&al2006,Pradhan&al2010}) to -4.5 for LFBM4. Interestingly, both the DFBM and the LFBM8 exhibit a similar critical exponent.
Increasing the spatial correlation of fiber strength (CL) changes the statistical behavior of fiber failure avalanche sizes. We find that while avalanches remain power-law distributed, the $b$-exponent for relative avalanche size larger than $10^{-5}$ decreases with increasing CL (Fig. \ref{SFD}). This dependency could be attributed to the increased frequency of failure cascades resulting from localization associated with CL and load sharing rule. 
To our knowledge, this is the first study that provides evidence for a change in critical exponent of avalanche size-frequency distribution when considering spatial correlation of the mechanical properties of a complex and heterogeneous system.
This behavior was more pronounced for LFBM4 than for LFBM8, suggesting that changes in the avalanche SFD are more sensitive to short-range interactions than to long-range interactions.

\begin{figure}

\includegraphics[width=8.6cm]{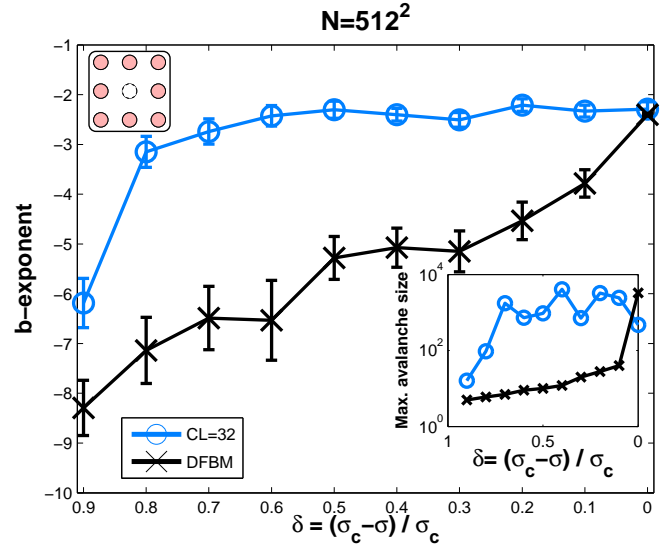}

\caption{\label{b_maxav} The evolution of the $b$-exponent (inset: the evolution of  maximum avalanche size) as a function of proximity to failure parameter $\delta$ for LFBM8 with ensemble average of 100 realizations to ensure good statistical representativity.}
\end{figure}
As LFBM8 exhibits a shift from progressive diffuse damage behavior (ductile-like rupture) to abrupt single crack growth behavior (brittle-like rupture) when increasing CL, this suggests that the same natural material, composed of similar mechanical components could exhibit vastly different global failure modes depending on the structural organization of its material properties.  

In order to characterize how fiber failure event size distribution evolves toward global failure, we calculated $b$-exponent of SFD for successive bins of relative stress to failure parameter $\delta$ (with a value of $\delta = 0$ marking failure). As the simulations were realized under stress controlled conditions, 
$\delta=\frac{\sigma_c-\sigma}{\sigma_c}$, $\sigma_c$ marking the stress at the critical point. 
To ensure reliable statistical representation we stacked 100 simulations for each configurations (DFBM, LFBM4 and LFBM8). The evolution of $b$-exponent when approaching final rupture is shown in Fig. \ref{b_maxav}. The simulated $b$-exponent increases as the system  approaches catastrophic rupture for ductile-like progressive rupture (DFBM and LFBM8 with CL=1), which may provide a useful tool for assessing the mechanical state  of a failure sensitive slope of geological structure, as \cite{Amitrano2012} already pointed out. With increasing CL, no trend in $b$-exponent was found, suggesting that the likelihood for acoustic or microseismic-based early detection of imminent failure is relatively limited based on the use of SFD in brittle materials. Similarly the evolution of maximum avalanche size was studied especially towards global failure (Fig. \ref{b_maxav}). The maximum avalanche size seems to increase when approaching global failure in ductile-like case, whereas for brittle-like failure (with large CL) avalanche size shows a peak and even a decrease before global failure, suggesting that a stagnant or even decrease of acoustic emission activity is expected to take place before a brittle-like failure.

\begin{figure}

\includegraphics[width=8.6cm]{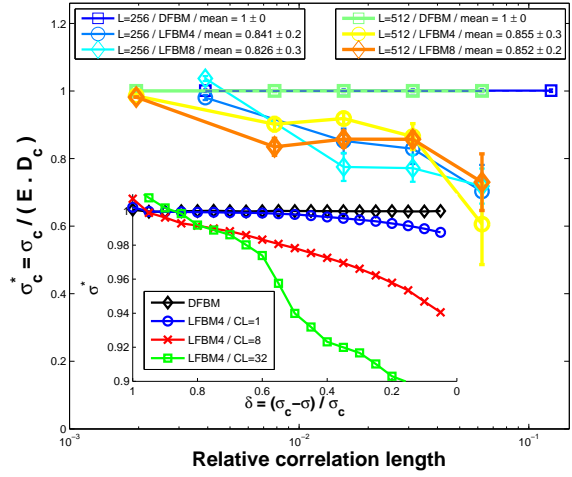}

\caption{\label{rupturecriterion} Rupture criterion and mean values for DFBM (square), LFBM4 (circle)  and LFBM8 (diamand) for N=256$\rm ^2$ and N=512$\rm ^2$ as a  function of relative correlation length, each point representing mean of 100 realizations. 
Inset: Evolution of $\sigma^{\star}$ (averaged over 100 realisations) as a function of the control parameter $\delta$.}
\end{figure}

We examined the total elastic energy radiated from failed fibers in comparaison with the total elastic energy stored in the FBM. The motivation was to identify a dimensionless invariant metric (across CL values and redistribution rules) that could characterize conditions for global failure in a fashion similar to a Griffith-like \cite{Griffith1920} crack propagation criterion. While the resulting values of the metric appears relatively constant, the results show large scattering (the ratio of standard deviation to the mean of about 30$\%$). These results prompted us to consider a different dimensionless metric that is estimated entirely from macroscopic parameters expressed as $\frac{\sigma_c}{E \times D_c}$ that appears to be invariant with only 10$\%$ scatter through all redistribution rules (DFBM, LFBM4 and LFBM8) and CL (Fig. \ref{rupturecriterion}). Note that this dimensionless criterion is not sensitive to the global failure mode (brittle-like or ductile-like) nor to the CL (i.e., the spatial heterogeneities inherently present in geomaterials).
This suggests that the global failure of complex materials is determined by a combination of both load and damage state independent of details of spatial correlation of mechanical properties of failure mode. 
This key result opens new perspectives for studying gravity-driven failures, as it suggests that the relevant parameter to investigate rupture in heterogeneous media is no longer the effective stress but a "damage weighted stress"  $\sigma^{\star}_c$ accounting also for damage accumulation and expressed as
$\sigma^{\star}_c~\sim~\frac{\sigma_c}{E \cdot D_c}~\sim~\frac{\sigma^{tot}_c}{E \cdot (1-D_c)~\cdot~D_c}$
 where $\sigma^{tot}_c$ is the total stress at rupture and $D_c$ the damage at global failure. 
Simulation results suggest that the value of the metric $\sigma^{\star}=\frac{\sigma^{tot}}{E \cdot (1-D)D}$ decreases toward its critical value (marking global failure). The decrease in this metric is more pronounced for larger values of spatial correlation length (Fig. \ref{rupturecriterion}).
 In other words, introducing additional damage may destabilize a low-damaged material more than an already highly-damaged geomaterial. Estimating the evolution of this metric could potentially serve as an indicator for imminence of failure of slopes and other geologic structures, especially for the case of brittle rupture (that otherwise provides limited precursory events).
Moreover, as  $\delta \sigma^{\star}(\sigma,D)~=~\frac{\partial \sigma^{\star}}{\partial \sigma}\delta \sigma~+~\frac{\partial \sigma^{\star}}{\partial D}\delta D~=~\frac{1}{(1-D)D}\delta \sigma~+~\frac{(2D-1)\sigma}{(1-D)^2~D^2}\delta D$, the relative impact on global stability of a change in $\sigma$ or $D$  can be investigated separately. It appears that $\frac{1}{(1-D)D} < \frac{(2D-1)\sigma}{(1-D)^2~D^2}~\;~~\forall 0<D<0.5,~~ \forall \sigma>0$, meaning that a change in internal damage has potentially more destabilizing effects than a change in external loading. 
 In other words, both internal and external effects (respectively, internal damage and external loading) contribute to a gravity-driven instability in a natural medium (soil, rock, ice, snow,...), with internal damage playing a dominant role.

Of a practical importance is the fact that this criterion is independent of spatial organisation of mechanical properties, and heterogeneities are not required to be characterized explicitly.
Estimating total stress applied on the system would enable to evaluate the critical damage leading to the global rupture. 
As the internal response of the system plays a central role in the global failure, this study points to the importance of characterizing internal damage and its evolution for proper assessment of slope stability. Although, from a practical point of view, the "damage" state of natural geo-materials is not accessibly, evidence suggests potential links between evolution of internal damage and acoustic emissions \cite{Michlmayr&al2012}. Hence the monitoring of unstable slope with acoustic sensors seems to be promising way to assess evolution of damage with time. Nevertheless, further investigations are needed, especially on the healing phenomenon that may reduce internal damage, a phenomenon not yet considered by the model.

{\it Conclusions.}- 
The failure characteristics of mechanical analogs of natural systems represented by various fiber bundle models (FBM) were systematically evaluated. The focus was on the roles of spatial correlation and load redistribution rules on failure statistics and global failure of the FBM.   
 Results indicate that FBM failure mode varies dramatically with increasing correlation length and localized load sharing rules. Systems with similar composition of mechanical elements exhibit a dramatic transition from ductile and diffuse damage for global load sharing  (DFBM) to brittle single failure for correlated and local load sharing (LFBM). These changes in mechanical responses also affect the statistical properties of fiber failure avalanches (micro-cracks) activity preceding rupture and sought after in various early warning scenarios \citep{Amitrano2012}. While diffuse damage behavior exhibits clear precursory signals (such as increased seismic activity prior to global failure), brittle failure occurs abruptly with only few precursors. The results suggest that heterogeneous slopes with limited spatial correlations (either in terms of soil type or root reinforcement) are likely to behave as ductile materials producing ample precursory events, whereas drier and well-cemented slopes with strong root reinforcement are likely to fail like brittle materials with limited early warning. 
Although increasing spatial correlations of mechanical properties promotes abrupt ruptures at lower external load, we obtained an "universal" global failure criterion based on macroscopic properties which is independent of the rupture mode, stress redistribution rules, or the spatial organisation of mechanical properties. This metric that considers the combined role of external load and cumulative damage provides a means for evaluating imminence of failure of heterogeneous materials without resolving details of the heterogeneity. 
This study provides new insights that are potentially useful for understanding landslide triggering and points out the importance of spatial organization of heterogeneities on the failure behavior of complex geomaterials.

\textit{Acknowledgments} -  
This work was supported by the Competence Center Environment and Sustainability of the ETH Domain (CCES) in the research project TRAMM (Triggering of Rapid Mass Movements in Steep terrain).

\end{document}